\documentclass[aps,prb,twocolumn,a4paper,groupedaddress,twocolumn,showkeys,showpacs]{revtex4}
\usepackage{graphicx}
\usepackage{tabularx}
\usepackage{amsmath}
\usepackage{amsfonts}
\usepackage{amssymb}
\usepackage{natbib}
\begin{document}
\title{Comment on ``Enhanced transmission of light through a gold film due to excitation of standing surface-plasmon Bloch waves"}
\date{\today}
\author{J. Weiner}\affiliation{IRSAMC/LCAR\\
Universit\'e Paul Sabatier, 118 route de Narbonne,\\31062
Toulouse, France\\and\\IFSC/CePOF\\ Universidade de S\~ao Paulo, Avenida Trabalhador S\~ao-calense, 400-CEP 13566-590\\ S\~ao Carlos SP, Brazil}
\email{jweiner@irsamc.ups-tlse.fr}

\begin{abstract}
The purpose of this Comment is first to correct a misapprehension of the role played by composite wave diffraction on surface-wave generation at subwavelength structures and second to point out that periodic Bloch structures are unnecessary for the efficient production of the surface plasmon polariton (SPP) guided mode either as traveling or standing waves.  Guided surface waves originate from simple slit or groove edges illuminated under normal incidence, and one-dimensional (1-D) surface cavities from these standing waves are easily realized.
\end{abstract}
\maketitle
The authors of Ref.\,\onlinecite{SH07}, report enhanced transmission of light through a thin gold film in a T-shaped structure acting as a Fabry-P\'erot resonator.  They interpret these results as showing that they ``strongly contradict" the composite diffracted evanescent wave (CDEW) model\cite{LT04} and cite an experimental study on subwavelength single slit-groove structures\cite{GAVa06} as evidence supporting the CDEW model.  This Comment points out that the situation is not quite that simple.  The purpose of two experimental studies \cite{GAVa06,GAVb06}, the first cited in Ref.\,\onlinecite{SH07}, and a second, complementary set of measurements \cite{GAVb06}, was to test two key predictions of the CDEW model: (1) surface wave amplitude behavior with distance from the originating structure and (2) the phase of the surface wave relative to the phase of the light wave incident on this structure.  These studies showed that the surface wave amplitude indeed falls off rapidly within a ``near-zone" of a few wavelengths distance from the originating groove edge, but that after the initial fall-off the surface wave continues to propagate with near-constant amplitude.  This ``far-zone" behavior suggested population of the SPP guided wave mode, but the measured surface index of refraction $n_{\mathrm{surf}}=1.04\pm 0.01$ did not correspond to the expected SPP index $n_{\mathrm{SPP}}=1.015$. The phase behavior measured in Ref.\,\onlinecite{GAVb06} did show a phase shift close to $\pi/2$ consistent with the CDEW model prediction.  These tests, therefore, presented evidence consistent with diffraction in the near-zone but also confirmed the population of a long-lived surface wave, not predicted by CDEW.  In a follow-up study\cite{GAW07} the anomalous surface index of refraction measurement was found to be due to the transient behavior of the surface wave in the near zone.  As the surface wave evolves from the near-zone to the far-zone it asymptotically approaches the expected SPP guided mode wavelength.  This behavior implies that in the transient near-zone the surface wave is indeed a composite, consisting of many surface modes including the bound SPP.  But surface modes other than the SPP rapidly damp within a few optical cycles into dissipative phonon or radiative channels, consistent with a Drude model permittivity for real metals.  This picture has been confirmed in another study of surface waves in slit-groove structures on gold films\cite{KGA07}.

The physics that emerges from these experiments is that composite diffraction and bound SPPs are not mutually exclusive phenomena.  The sharp edges of a subwavelength groove or slit diffract the normal-incident source waves into surface modes, all of which damp rapidly except for the bound SPP.  This process demonstrates remarkable efficiency with about 30-40\% of the initial surface wave amplitude evolving to the asymptotic SPP, although the details of how this evolution occurs are not yet well understood.  The relevant point to emphasize here, however, is that the oft-stated belief that grating or prism coupling is required to supply the missing momentum for efficient generation of the SPP mode is simply not true.  The momentum spectrum associated with a grating structure of period $a$ will be narrowly peaked at $2\pi/a$, while the momentum distribution of a subwavelength slit of width $w$ will be broad, essentially its Fourier transform, $\simeq 2\pi/w$, and centered at $2\pi/\lambda_0$ with $\lambda_0$ the normal-incident wavelength.  The use of a grating structure will produce ``Bloch modes" on the surface and within the skin depth of the metal, but there appears to be no reason to invoke structural periodicity as a \emph{necessary} feature for the generation of SPPs and no need to distinguish ``Bloch" surface waves from ``regular" surface waves.
\begin{figure*}
\begin{minipage}[t]{0.75\linewidth}
\includegraphics[width=\columnwidth]{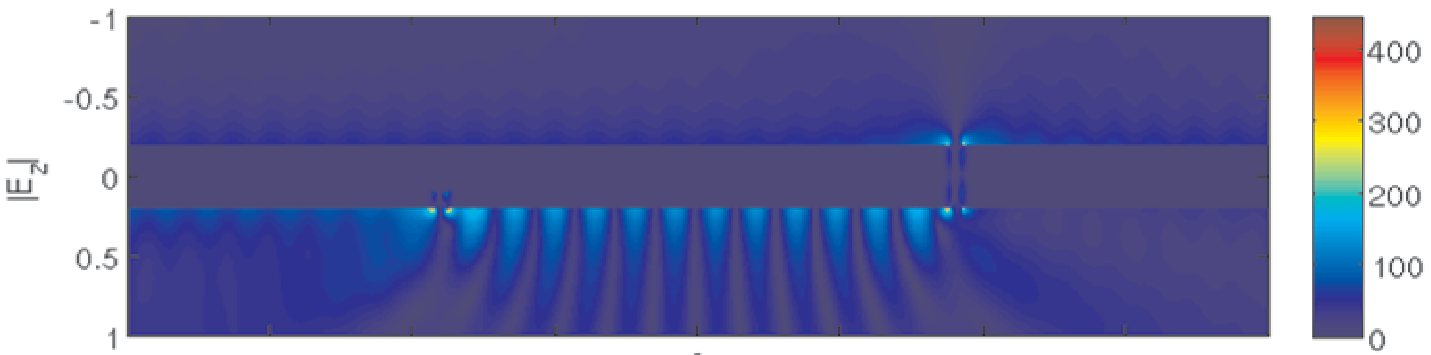}\caption{(color online) FDTD calculation of $E_z$ amplitude as a function of $z$ (perpendicular distance from the slit-groove reference plane) and $y$ (transverse distance along the slit-groove reference plane.  The center-to-center distance between the slit and groove is 3.6~$\mu$m.  Slit and groove are both milled 100~nm wide; the groove depth is 100 nm, and the silver layer is 400 nm thick.  The incident light has a free-space wavelength $\lambda_0=852$~nm. This plot is an unpublished result from Y. Xie and M. Mansuripur, shown here with permission.}\label{Fig-E-field-plot}
\end{minipage}\hfill
\begin{minipage}[t]{0.75\linewidth}
\includegraphics[width=\columnwidth]{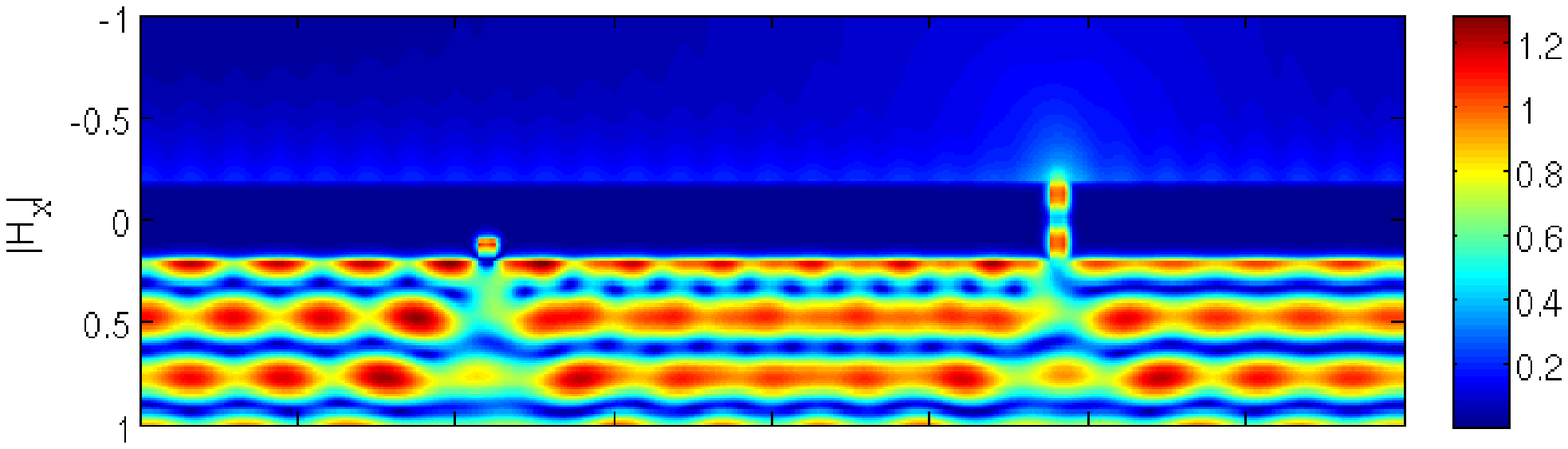}\caption{(color online) FDTD calculations of $H_x$ amplitude as a function of $z$ and $y$ for same structures as in Fig.\,\ref{Fig-E-field-plot}.  The $H_x$ component of the magnetic field is aligned with the long axes of slit and groove.  This plot is an unpublished result from Y. Xie and M. Mansuripur, shown here with permission.}\label{Fig-H-field-plot}
\end{minipage}
\end{figure*}
The authors of Ref.\,\onlinecite{SH07} launch SPPs into a smooth 1-D surface cavity from an adjacent periodically structured region, but the studies on single slits and grooves previously cited suggest that they would have obtained similar results (with different cavity Q-factors) by just using a subwavelength groove launcher.  In fact the far-field fringes measured in Refs.\,\onlinecite{GAVa06,GAVb06,GAW07,KGA07} result from the interference between a directly transmitted reference wave and the partially transmitted standing wave of a 1-D cavity formed by the slit and groove.  Figures \ref{Fig-E-field-plot} and \ref{Fig-H-field-plot} show finite-difference-time-domain (FDTD) simulations of the experiment in Ref.\,\onlinecite{GAVa06}. A train of plane waves with $\lambda_0=852$~nm, incident from below, illuminates the slit-groove structure.  The structure is a 400~nm layer of silver metal with a 100~nm wide slit milled through and a 100~nm wide groove milled to a depth of 100~nm.  The slit and groove are separated by a center-to-center distance of 3.6~$\mu$m.  Figure \ref{Fig-E-field-plot} plots the amplitude of the electric field component $E_z$ perpendicular to the surface plane as a function of $z$ and $y$, the distance along the groove-slit reference plane.  Figure \ref{Fig-H-field-plot} plots the magnetic field component $H_x$ (the component parallel to the long axis of the slit and groove) as function of the vertical and transverse dimensions of the structure.  Concentration of $E_z$ amplitude corresponds to surface charge concentration, and the ''cavity" standing wave pattern on the surface between the slit and groove is evident in Fig.\,\ref{Fig-E-field-plot}.  Another significant feature of Fig.\,\ref{Fig-E-field-plot} are the ``hot spots" around the corners of the slit and groove on the lower surface as well as the appearance of hot-spot charge concentrations around the corners of the slit on the upper surface.  These hot spots are not numerical artifacts of the FDTD simulation; they reveal the presence of localized oscillating charge that are the source of evanescent and propagating modes on the lower and upper surfaces of the structure \cite{XZM05,XZM06,LMW07}. It is also worth noting from inspection of Fig.\,\ref{Fig-H-field-plot} that the groove and slit themselves constitute optical cavities.  The quality factor of the 1-D surface cavity and the transmission from the incident (lower) side of the structure to the transmitted (upper) side can be optimized by a judicious choice of groove depth and metal layer thickness.  This surface-cavity/slit-cavity coupling controls the physics of optical transmission through subwavelength metallic slit arrays.


\begin{thebibliography}{99}
\bibitem{SH07}I. I. Smolyaninov and Y.-J. Hung, Phys. Rev. B \textbf{75}, 033411 (2007).
\bibitem{LT04}H. J. Lezec and T. Thio, Opt. Express \textbf{12}, 3629 (2004).
\bibitem{GAVa06}G. Gay, O. Alloschery, B. Viaris de Lesegno, C. O'Dwyer, J. Weiner, and H. J. Lezec, Nat. Phys. \textbf{2}, 262 (2006).
\bibitem{GAVb06}G. Gay, O. Alloschery, B. Viaris de Lesegno, J. Weiner, and H. J. Lezec, Phys. Rev. Lett. \textbf{96}, 213901 (2006).
\bibitem{GAW07}G. Gay, O. Alloschery, J. Weiner, H. J. Lezec, C. O'Dwyer, M. Sukharev, and T. Seideman, Phys. Rev. E \textbf{75}, 016612 (2007).
\bibitem{KGA07}F. Kalkum, G. Gay, O. Alloschery, J. Weiner, H. J. Lezec, Y. Xie, and M. Mansuripur, Opt. Express \textbf{15}, 2613 (2007).
\bibitem{XZM05}Y. Xie, A. R. Zakharian, J. V. Moloney, and M. Mansuripur, Opt. Express \textbf{13}, 4485 (2005).
\bibitem{XZM06}Y. Xie, A. R. Zakharian, J. V. Moloney, and M. Mansuripur, Opt. Express \textbf{14}, 6400 (2006).
\bibitem{LMW07}G. L\'ev\^eque, O. J. F. Martin, and J. Weiner, arXiv.0704.0703v1 (2007).
\end{thebibliography}
\end{document}